\documentclass{aa}
\usepackage{graphicx}
\usepackage{txfonts}
\usepackage{natbib}
\usepackage{ulem}
      \def\new#1 {{\bf #1 }}
      \def\cut#1 {\sout{#1} }
      \def\new#1 {#1 }
      \def\cut#1 { }
      \def\revised#1 { { #1 }}
\bibpunct{(}{)}{;}{a}{}{,}
\newcommand{\ngc}{NGC6334(I)}
\newcommand{\gthree}{G327.3-0.6}
\newcommand{\sotw}{\mbox{SO$_2$}}
\newcommand{\nthp}{\mbox{N$_2$H$^+$}}
\begin{document}
\title{Submillimeter spectroscopy of southern hot cores:\\ NGC6334(I) and G327.3-0.6}


\author{P. Schilke\inst{1} \and C. Comito\inst{1} \and S. Thorwirth\inst{1} \and F. Wyrowski\inst{1} \and K.M.
  Menten\inst{1} \and R. G\"usten\inst{1}  \and P. Bergman\inst{2} \and L.-\AA. Nyman\inst{2} }

   \offprints{P. Schilke}

   \institute{Max-Planck-Institut f\"ur Radioastronomie, Auf dem H\"ugel 69,
     53121 Bonn, Germany\\
     \email{[schilke, ccomito, sthorwirth, wyrowski, kmenten,
       guesten]@mpifr-bonn.mpg.de} \and European Southern Observatory, Alonso
     de Cordova 3107,
     Vitacura, Casilla 19001, Santiago, Chile \\
\email{[pbergman, lnyman]@eso.org}
   }

   \date{Received April 2006, Accepted hopefully soon}


  \abstract
   {High-mass star-forming regions are known to have a rich molecular
   spectrum from many species.  Some of the very highly excited lines are
   emitted from very hot and dense gas close to the central object(s). }
   {The physics and chemistry of the inner cores of two high mass star forming
   regions, \ngc\ and \gthree,  shall be characterized. }
   {Submillimeter line surveys with the APEX telescope provide spectra which
   sample many molecular lines at high excitation stages.}
   {Partial spectral surveys were obtained, the lines were identified,
   physical parameters were determined through fitting of the spectra. }
 {Both sources show similar spectra \cut{than} \new{that are comparable to
     that of} the only other high mass star forming region ever surveyed
   \new{in this frequency range}, Orion-KL, but with an even higher line
   density.  Evidence for very compact, very hot sources is found.}

   \keywords{Astrochemistry, Line: Identification, ISM: Molecules
               }

   \maketitle
%

\section{Introduction}
Unbiased spectral line surveys are a unique way of characterizing astronomical
sources. However, they are very time consuming, and the data analysis is not
easy, so the number of line surveys has been very limited (see
\citealt{comito05} and references therein), both in numbers of sources (but
see \revised{\citealt{macdonald96, hatchell98, thompson99, thompson03}}) and
in frequency range covered.  There is only one source for which all
\new{sub-THz} atmospheric windows have been surveyed from the ground,
Orion-KL.  However, with the advent of more sensitive instruments with larger
bandwidths, and with more sophisticated analysis methods, the situation is
slowly changing, and even seemingly unlikely targets such as solar-mass
protostars \cut{are} \new{can be} targeted \citep{castets05}.  However, the
submillimeter spectral regime, shortward of 800~$\mu$m, is still relatively
unexplored, apart from Orion-KL.

At the location of APEX\footnote{This publication is based on data acquired
  with the Atacama Pathfinder EXperiment (APEX).  APEX is a collaboration
  between the Max-Planck-Institut f\"ur Radioastronomie, the European Southern
  Observatory, and the Onsala Space Observatory.} (G\"usten et al. 2006, this
volume), the site quality is so superior to any other site at which large
submillimeter telescopes reside, that exploring the submillimeter regime on a
more regular basis has become a possibility.  Hence, we have started
submillimeter line surveys toward two very promising southern high-mass star
forming regions (see Tab.~\ref{tab:sources} for a description), NGC6334(I) and
G327.3-0.6.  
\begin{table}[htbp]
  \centering
  \begin{tabular}{lrrccc}
\hline
    Object & $\alpha$ (J2000) & $\delta$ (J2000) & Dist. & Mass &
    Lum.\\
    & & & kpc & $M_\odot$ & $10^5 L_\odot$\\
\hline
 \ngc & 17:20:53.4 &  -35:47:01 & 1.7 & 200 & 2.6\\
 \gthree &  15:53:08.8 & -54:37:01 & 2.9 & 500 & 0.5-1.5\\
\hline

  \end{tabular}
  \caption{Parameters of sources observed. \revised{The luminosities and
    masses for \ngc\ and \gthree\ were taken from \citet{sandell00} and
    Wyrowski et al., (this volume), respectively.} }
  \label{tab:sources}
\end{table}
NGC6334(I) has been the target of a millimeter line survey
\citet{thorwirth03}, which showed a plethora of lines \new{from many
  molecules}.  Earlier observations \citep{bachiller90, mccutcheon00} already
proved that this source is a chemically rich hot core.  G327.3-0.6 had been
studied by \citet{bergman92}, and part of its molecular content has been
sampled by \citet{gibb00}, who found that complex organic molecules are more
abundant here than in any other Galactic object, including Orion-KL and
SgrB2(N). In this \emph{Letter} we give a first account of the kind of results
one can expect from submillimeter surveys of chemically rich sources.  More
observations are planned, which will give a more complete census of the
molecules present, and will allow a more rigorous determination of the
physical parameters.

\section{Observations}

\begin{figure*}[htbp]
  \centering
  \includegraphics[scale=0.3,angle=-90]{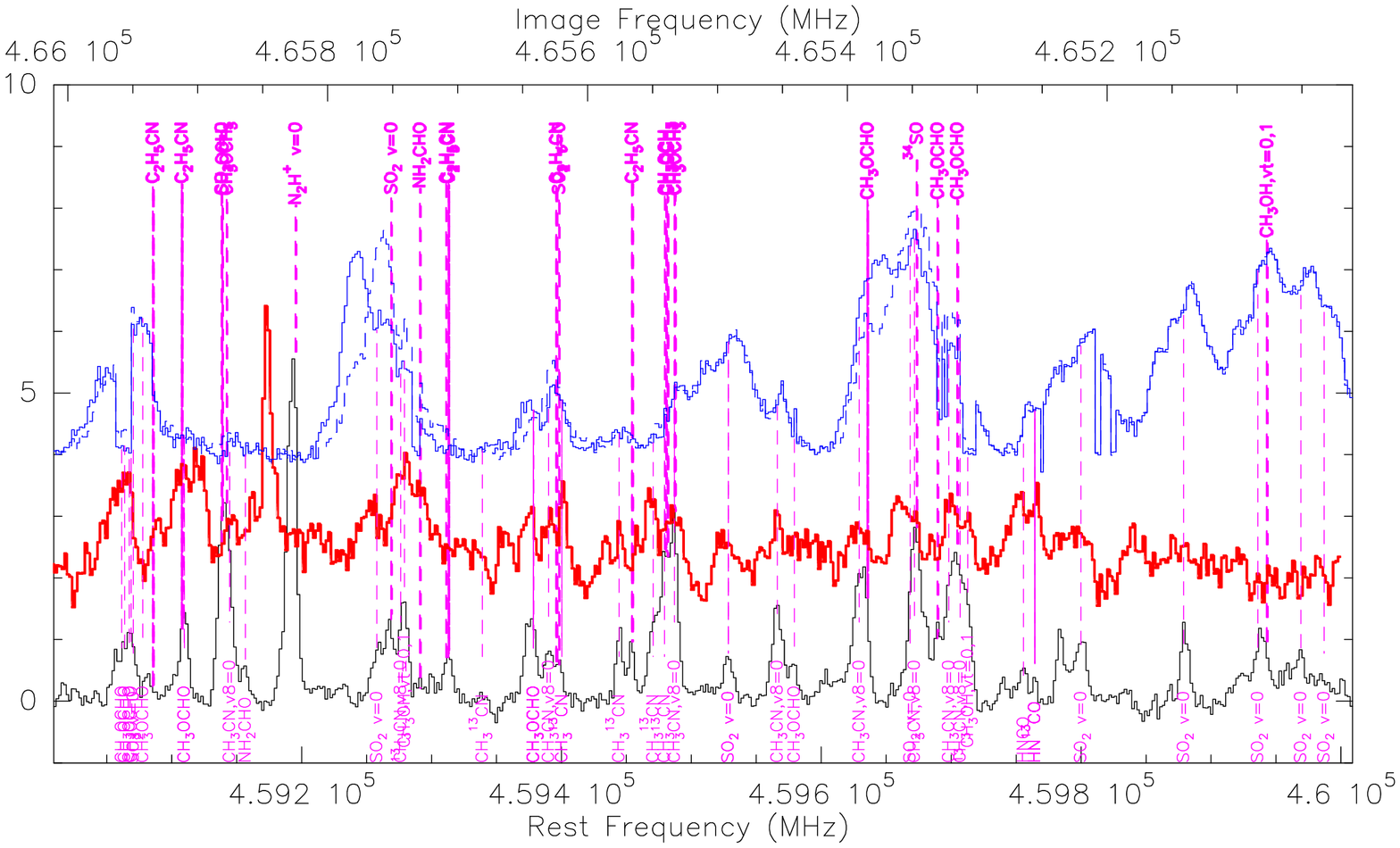}
  \includegraphics[scale=0.3,angle=-90]{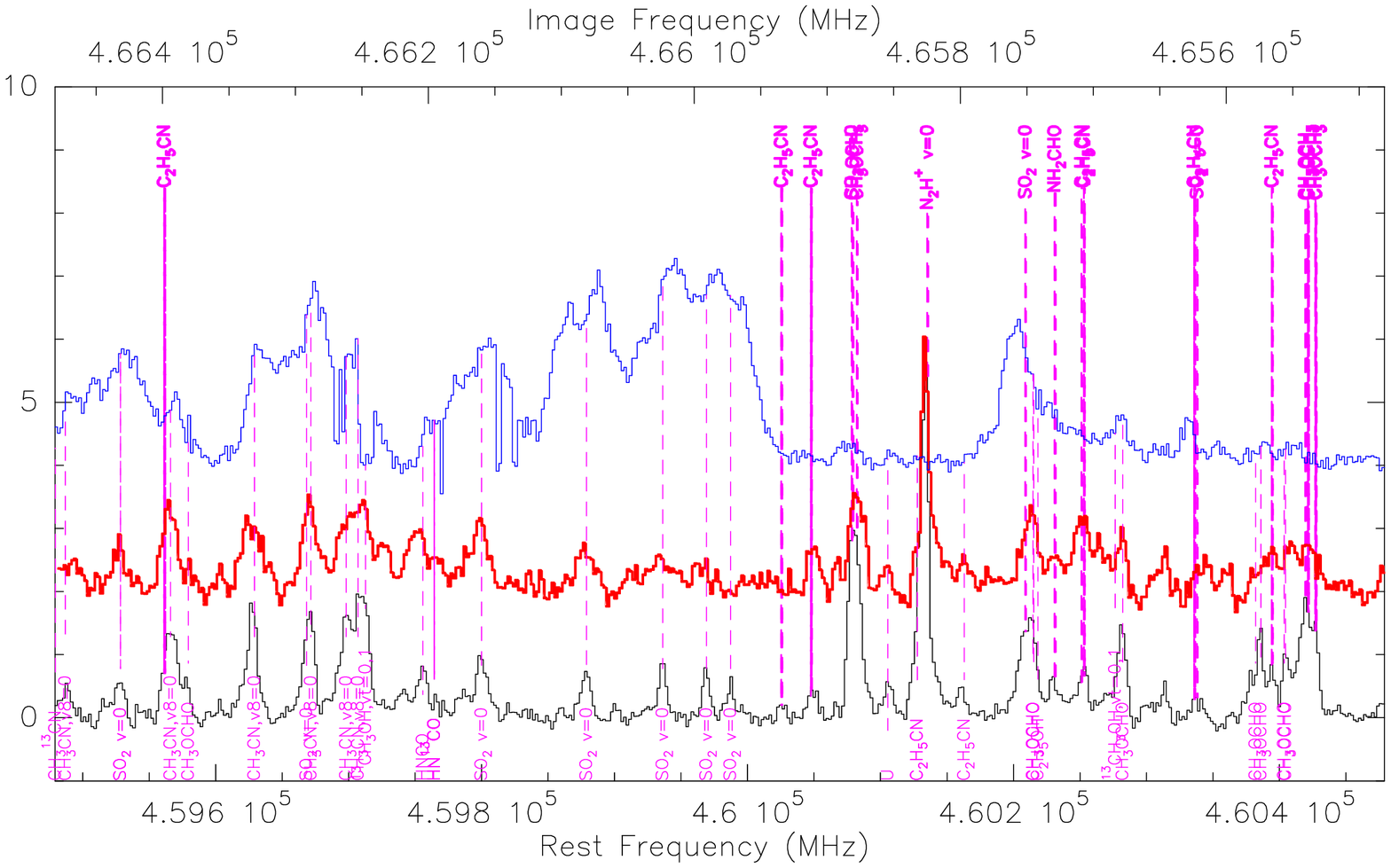}
  \includegraphics[scale=0.3,angle=-90]{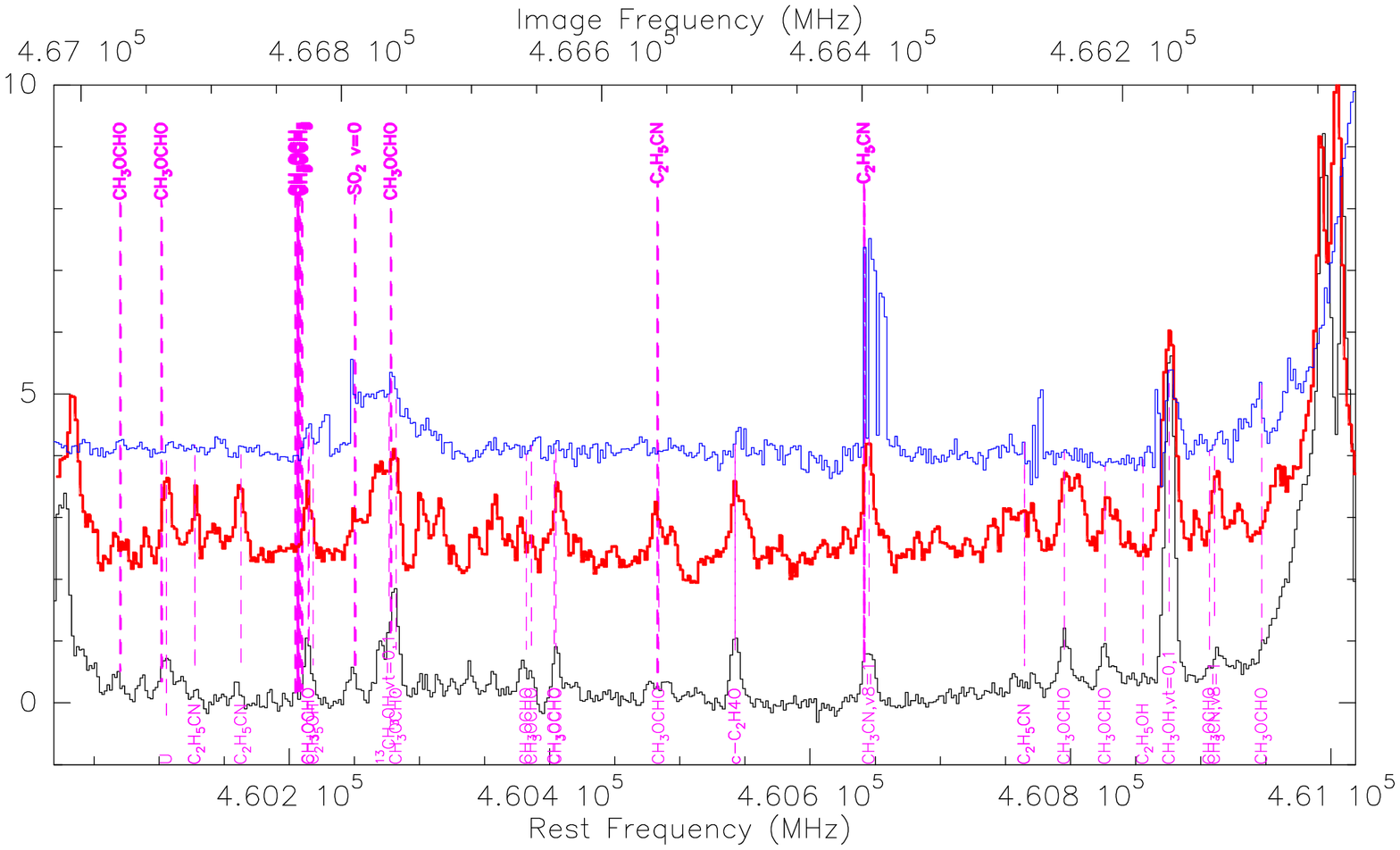}
  \includegraphics[scale=0.3,angle=-90]{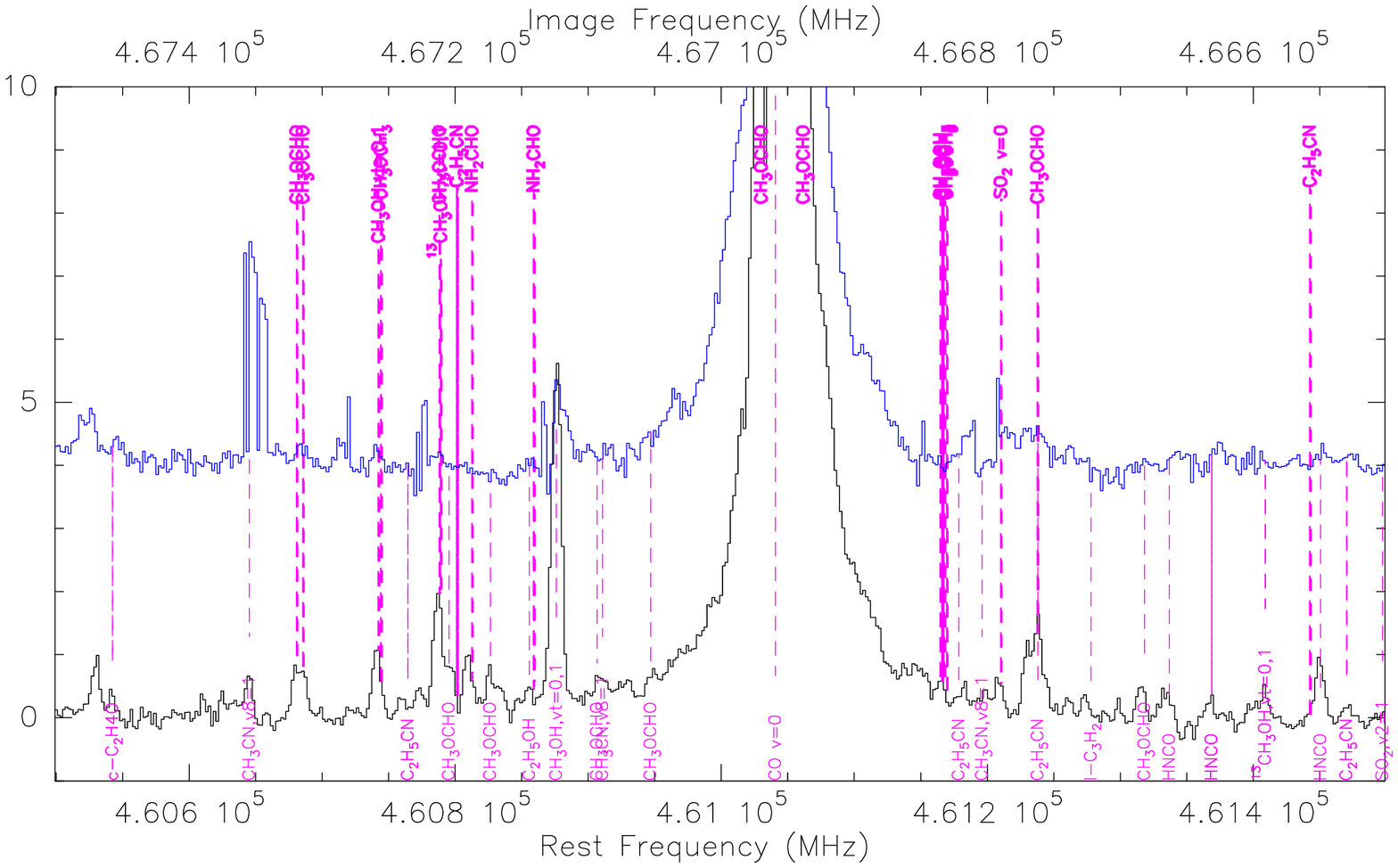}
  \includegraphics[scale=0.3,angle=-90]{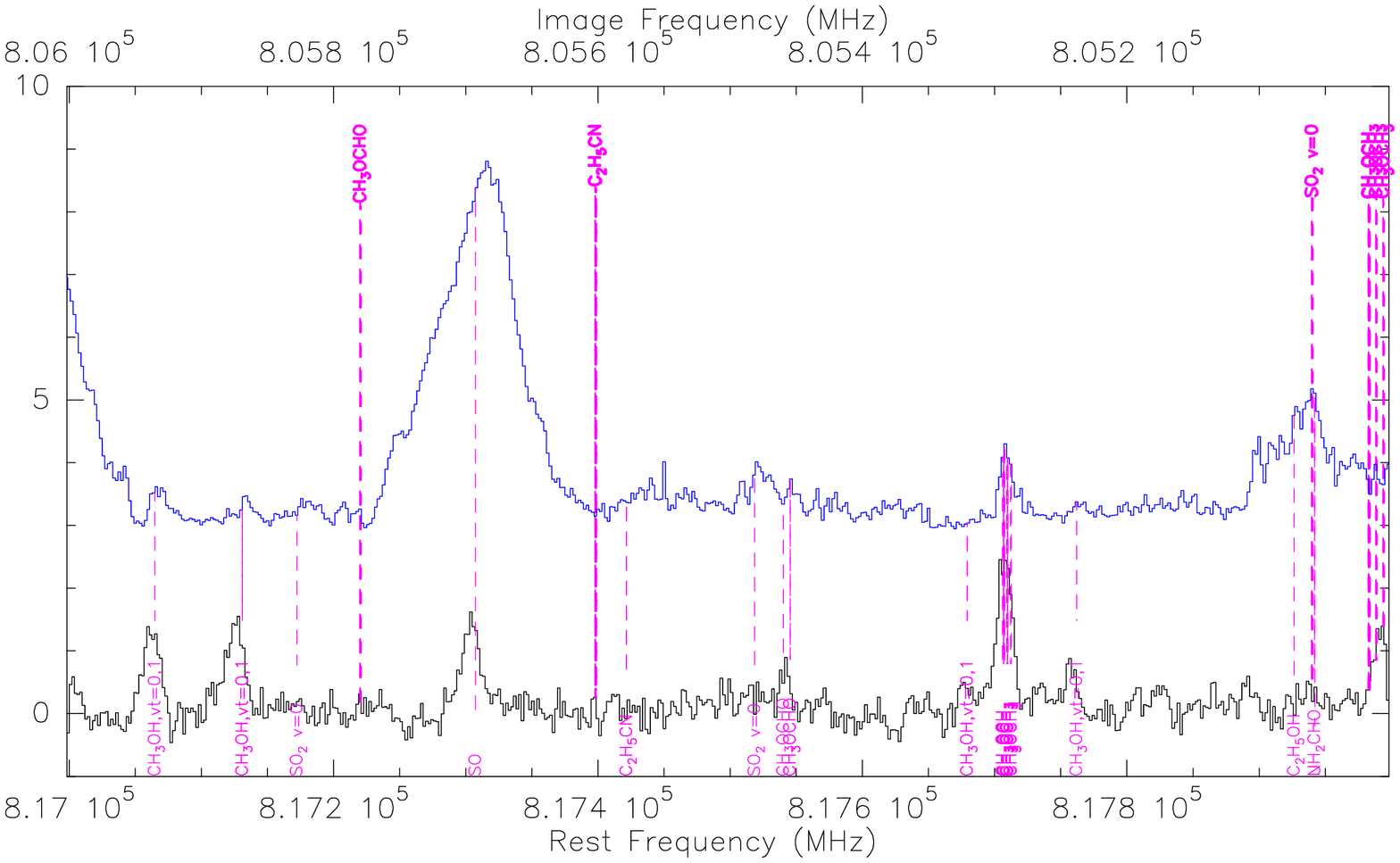}
  \includegraphics[scale=0.3,angle=-90]{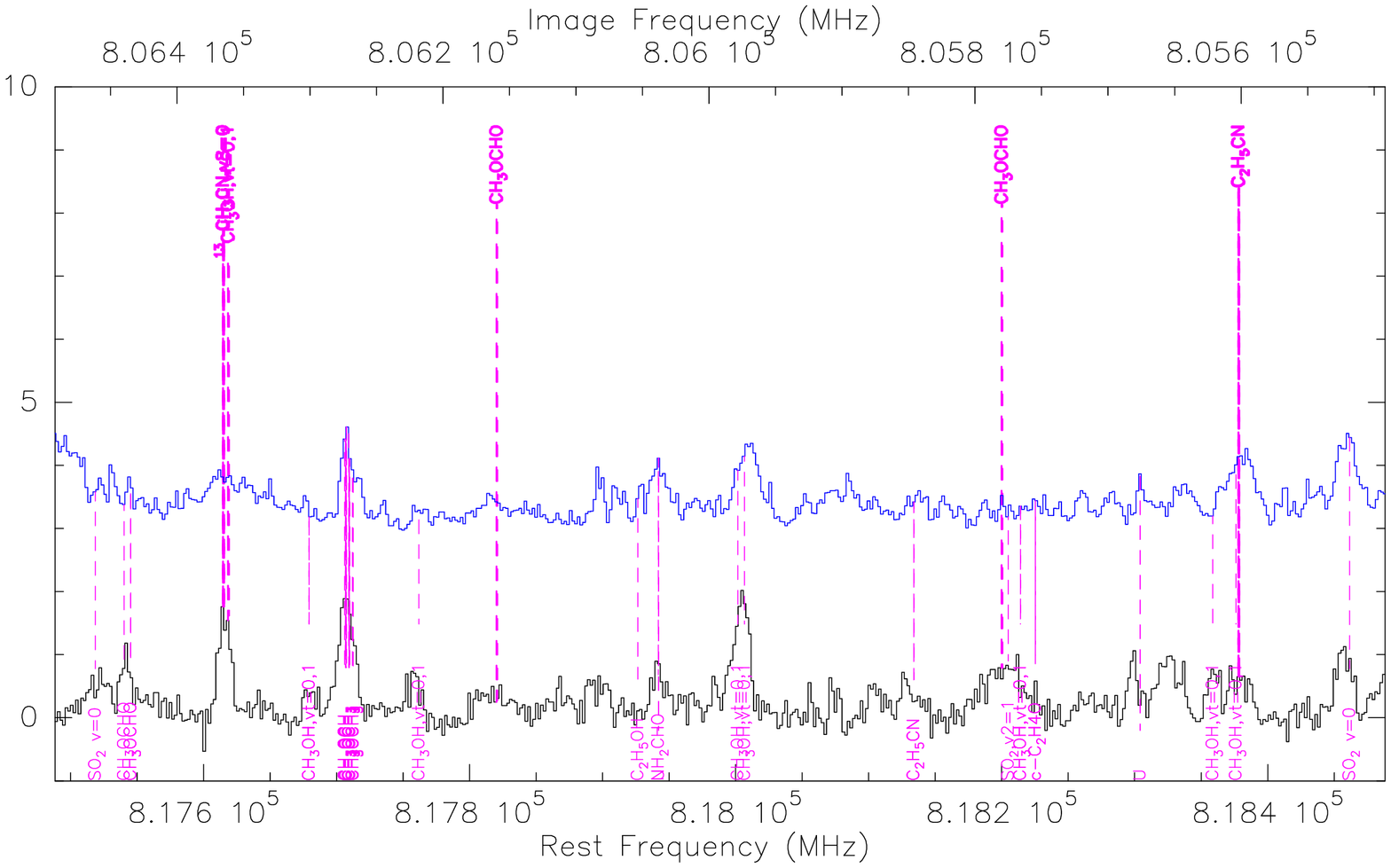}
  \includegraphics[scale=0.3,angle=-90]{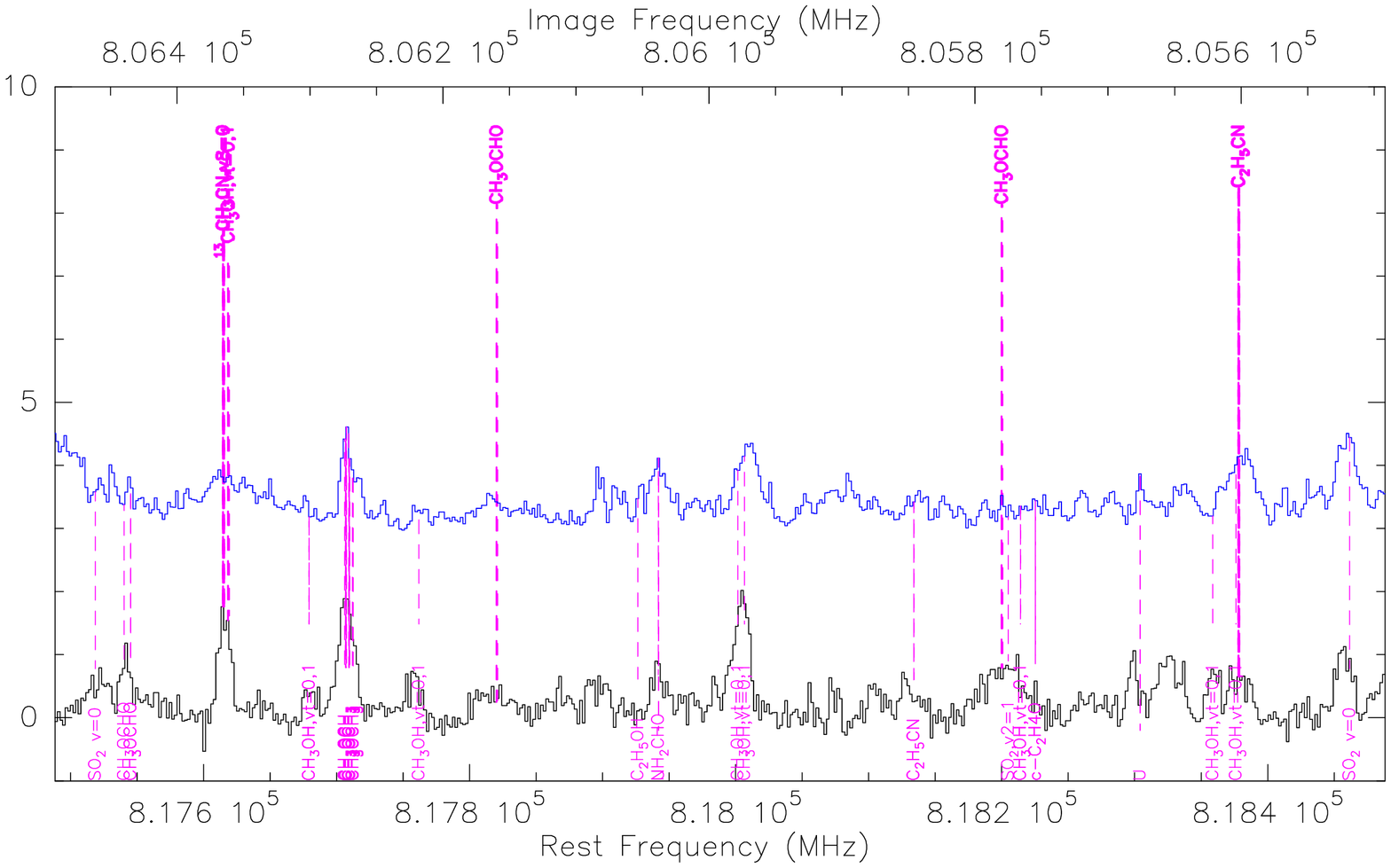}
  \includegraphics[scale=0.3,angle=-90]{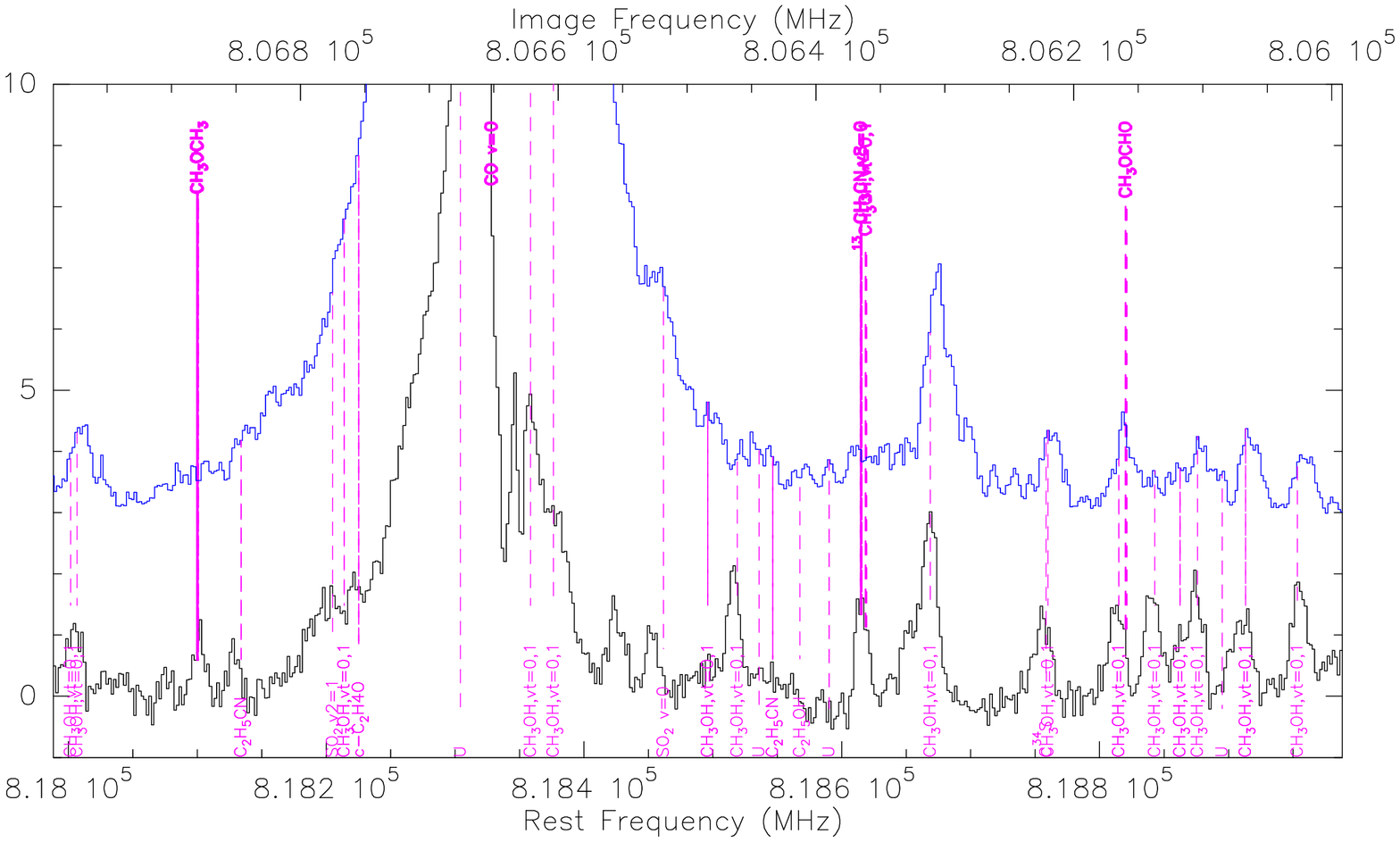}

  \caption{Spectra of the line survey toward NGC6334(I) (black, bottom) and
    G327.3-0.6 (red, center), compared to synthetic DSB spectra of Orion-KL
    (blue,top). In the first spectrum (top left), the NGC6334(I) and G327.3-0.6
    spectra were not taken with the same frequency setting, so the lines from
    the other sideband are somewhat shifted.  In this case, the dashed blue
    line is the synthetic Orion-KL DSB spectrum for the NGC6334(I) setting,
    the solid blue line is the Orion-KL DSB spectrum for the G327.3-0.6
    setting.  The Orion spectra have been scaled by factors 0.05 and 0.1 for
    the 460~GHz and 810~GHz bands, respectively.}
  \label{fig:spec}
\end{figure*}

The line survey was started in June 2005, using the FLASH instrument on APEX
(Heyminck et al. 2006, this volume) and the FFTS as backend (Klein et al.
2006, this volume).  For NGC6334(I), we had both the 460 and the 810~GHz bands
available, with beam sizes of 14$''$ and 8$''$, respectively; for G327.3-0.6,
we only took 460~GHz spectra due to technical reasons.  Following the
recommendations of \citet{comito02}, we sampled the spectra with a redundancy
of 2, using irregular sampling, i.e.\ at 1~GHz bandwidth, the frequency
separation between adjacent LO settings was about (but not exactly) 500~MHz.
Due to time constraints, only 2.5~GHz in NGC6334(I) was covered in
each band, and only 2~GHz in G327.3-0.6 (see Fig.~\ref{fig:spec} for spectra).
Pointing was performed on the continuum of the sources themselves.  For
\gthree, this resulted in an offset, because the nominal source position was
inaccurate (Wyrowski et al. 2006, this volume), but the pointing procedure
ensured that the observations were performed on the true source position.

\begin{figure*}[htbp]
  \centering
  \includegraphics[scale=0.3,angle=-90]{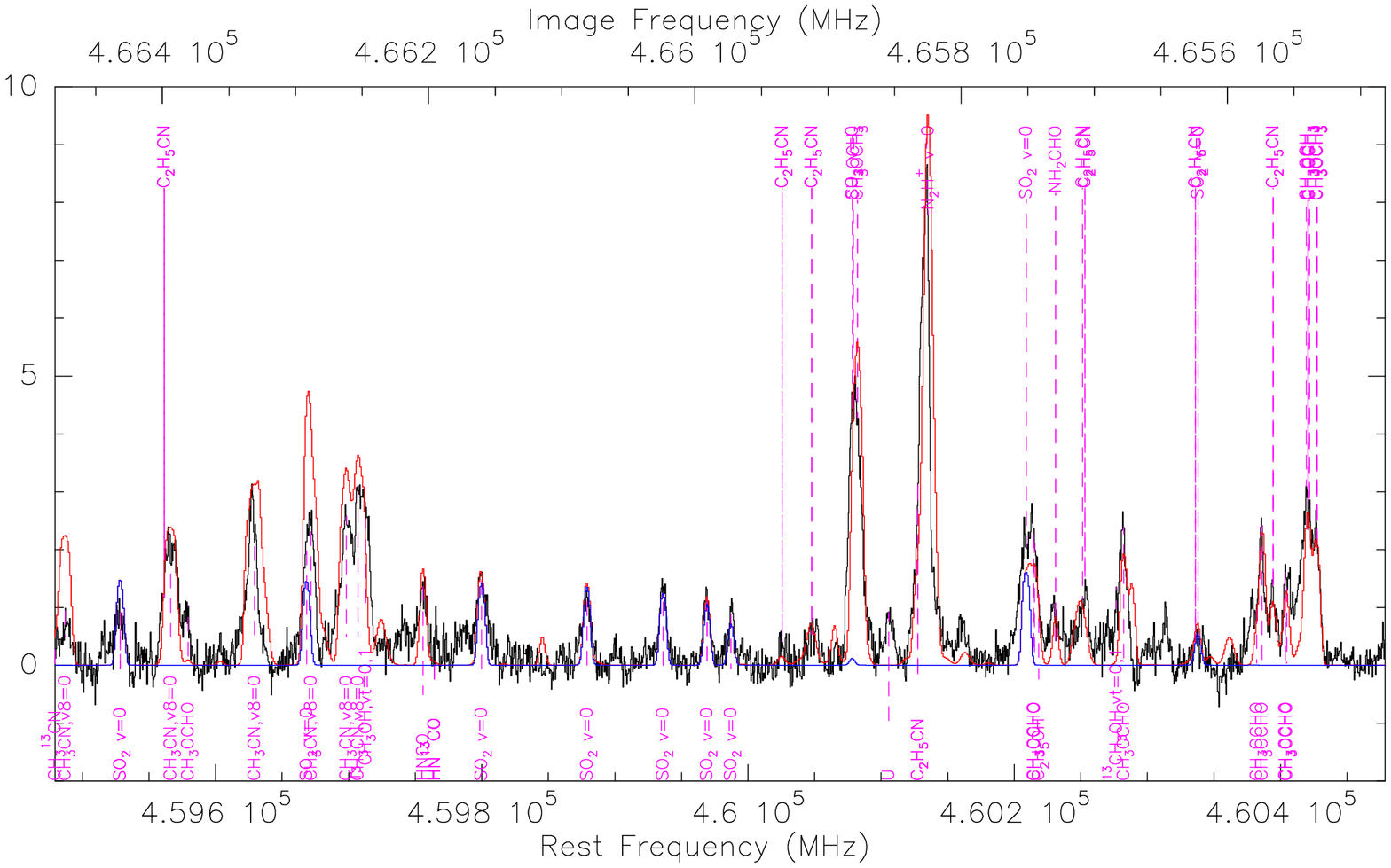}
  \includegraphics[scale=0.3,angle=-90]{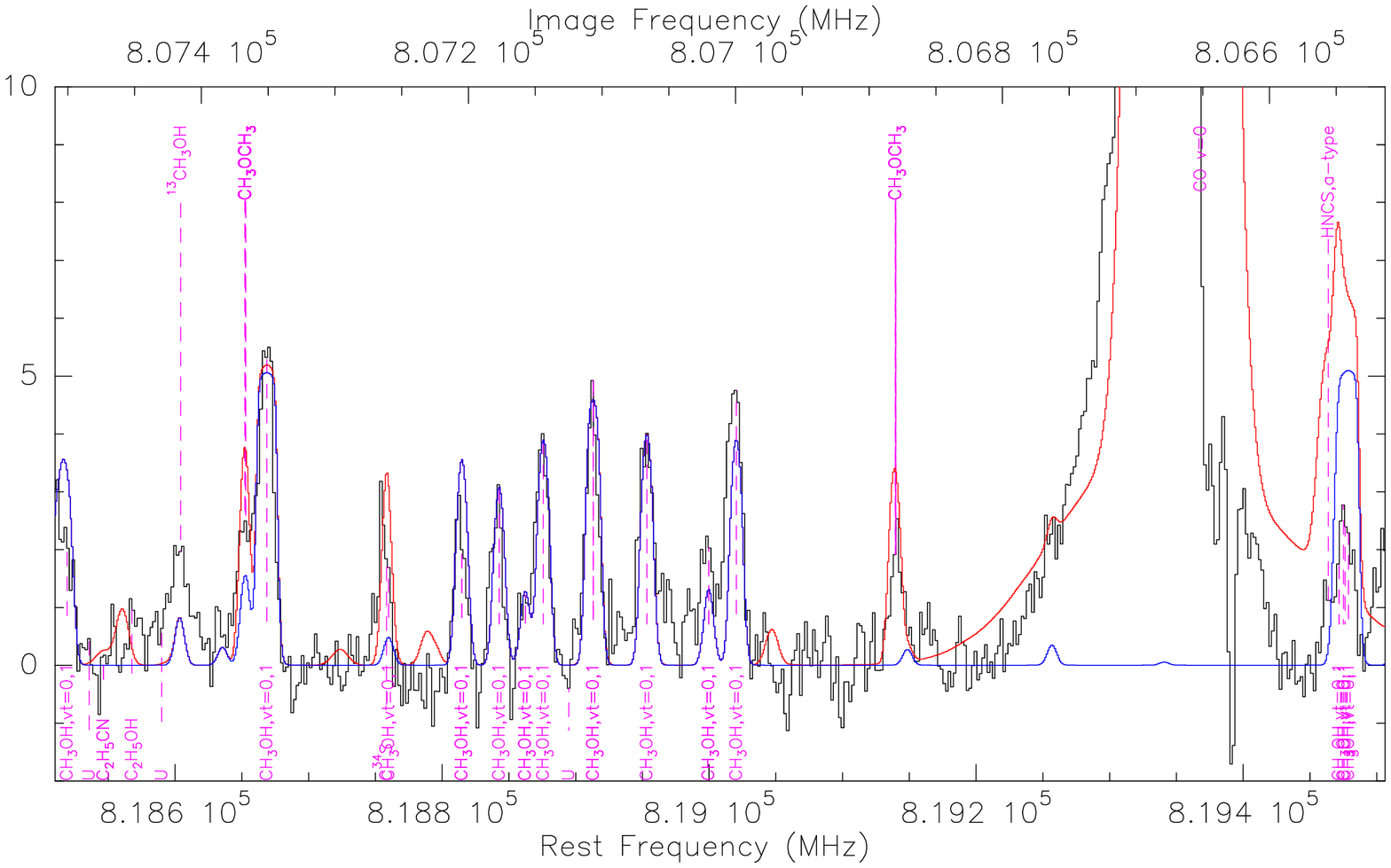}
  \caption{Spectra of the line survey toward NGC6334(I) (black) with model
  (red) overplotted. In blue, \sotw\ is shown in the left picture, in the
  right frame, blue depicts methanol.  }
  \label{fig:model}
\end{figure*}

\section{Analysis}
The \new{frequency coverage of the} dataset is not large enough for sideband
deconvolution, so our analysis was carried out on double side band (DSB)
spectra.  Since the frequency range surveyed is very small, we do not detect
lines from all molecules expected to be seen in these source, and for some
molecules the lines in our bands are not the best suited for temperature or
column density determination.  \cut{The results for column densities, column
  densities etc.\ we obtain are therefore have} \new{Our physical parameter
  determinations have therefore} to be regarded as preliminary in many cases,
and will be superseded once a larger data set is available.  Nonetheless, the
spectra of some molecules are well sampled, and our dataset allows an
assessment of particularly the high excitation regime, which dominate the
submillimeter spectra discussed here.

In Fig.~\ref{fig:spec} we show an overview of the data.  For comparison, we
have produced DSB spectra \revised{with the same LO setting as our data} of
Orion-KL in the same frequency ranges, \revised{using the published SSB survey
  data} of \citep{white03} and \citep{comito05} in the 460~GHz and 850~GHz
range, respectively. Note that the \cut{S/N} \new{signal-to-noise ratio} of
our data \new{matches that of} the Orion data, although the Orion lines are a
\cut{factor of 10-20 stronger} \new{between 10 and 20 times stronger}, because
Orion at 450~pc is much closer.  This is partly due to the better receivers we
employ, but mostly reflects the superior site quality.

At first glance, the spectra of \ngc\ and \gthree\ look very similar to each
other, although the lines of \gthree\ are somewhat broader than the lines
toward \ngc.  The line density seems to be higher in \gthree. The spectra
measured toward Orion-KL, on the other hand, look very different, as shown
particularly in Fig.~\ref{fig:spec}, top panels.  Closer inspection reveals
that the difference is mostly due to lines of sulphur bearing species, SO and
\sotw, which are known to be exceptionally strong and wide in this source.
Outside of these bands, Orion-KL shows markedly fewer features than the other
sources.  The absence of \nthp(5--4) in the Orion spectra can be accounted for
if we consider the difference in coupling to the beam by the three objects:
due to its proximity (450~pc), the angular size of Orion-KL is such that APEX
sees mostly the emission from the hot core, whereas toward \ngc\ and \gthree\ 
(1.7 and 2.9~kpc distance, respectively), the more quiescent gas around the
hot cores contributes significantly to the emission collected by the APEX
beam. However, most of the other lines which are absent in Orion are genuine
hot core lines, emphasizing the fact that Orion-KL's claim to fame is mostly
based on its close distance, but not to being truly exceptional.  In fact, it
is located on the low end of the luminosity and low spectrum of hot cores.

The analysis of the data was carried out with the \texttt{xclass} program
(discussed in \citealt{comito05}), which uses molecular data compiled from the
CDMS \citep{mueller01, mueller05} and JPL \citep{pickett98} databases and an
LTE model to produce synthetic spectra, which are then compared to the
observations.  The parameters defining the synthetic spectrum are, for each
molecular species: source size (to take beam coupling into account), rotation
temperature, column density, velocity width and velocity offset (with respect
to the systemic velocity of the object).  Several velocity components, which
are supposed to be non-interacting (i.e.\ the intensities add up
\new{linearly}) can be used, although in the present case we restricted the
model to two components, given the small amount of data. The source size is
degenerate with temperature in the case of completely optically thick lines,
and with column density for completely optically thin lines, in the sense that
degenerate parameters cannot be determined independently.  The advantages are
that, since all species are fitted simultaneously, blending is taken into
account, which otherwise presents a severe problem.  It is also possible to
fit molecules simultaneously with their isotopologues.

Error ranges for the various parameters can be estimated by a $\chi^2$
analysis, although this can only give errors within the assumptions of the
model (finite number of homogeneous components, LTE), but cannot assess errors
due to these assumptions.  On the other hand, densities in hot cores usually
are in a range which makes LTE a reasonable assumption (although this argument
is weaker for some of the very high excitation lines present in the spectra,
which do require either very high densities or IR pumping to be excited).
Most importantly, more sophisticated radiative transfer modeling would suffer
from missing collision rates for many of the species present here, and from
too many free parameters describing the source structure, which for the cores
in question is less well known than for northern, better studied cores.

The results in Tab.~\ref{tab:model} were derived as follows: the source sizes
were determined whenever possible (i.e.\ when some of the lines were optically
thick). In the remaining cases, the source was assumed to be extended, i.e.\ 
the beam averaged column density is given.  Similarly, we determined a value
the temperature whenever possible, otherwise a canonical value of 150~K was
assumed, except for \nthp, which is known to be absent in hot gas, and where
50~K were used.  Examples of the fits are shown in Fig.~\ref{fig:model}.
Without an error analysis, the results have to be consider preliminary, but a
more rigorous treatment of data analysis will be presented once larger ranges
have been surveyed.

\begin{table}[htbp]
  \centering
  \begin{tabular}{lrrrrl}
\hline
Line & Size & $T_{\rm ex}$ & $N$(mol) & $\Delta v$ & Comment\\
     & $''$ & K & cm$^{-2}$ & km s$^{-1}$ & \\
\hline
\multicolumn{6}{c}{\textbf{NGC6334(I)}}\\
\hline
c-C$_2$H$_4$O & ext & 150 & 4.0 (14)&  3.0 & 1\\
C$_2$H$_5$CN & ext & 150 & 2.0 (15) & 5.0 & 2\\
CH$_3$CN & 1.5 & 200 & 1.5 (17) & 4.0 & 3, 4\\
         & ext & 50 & 1.0 (15) & 4.0\\
CH$_3$OCH$_3$ & 1.2 & 200 & 2.0 (18) & 3.0 & 4\\
              & ext & 50 & 1.5 (16) & 5.0\\
CH$_3$OCHO & 1.5 & 200 & 1.0 (18) & 3.0 & 4\\
CH$_3$OH & 1.5 & 150 & 5.0 (18) & 3.0 & 4\\
         & ext & 50 & 3.0 (16) & 5.0 \\
CS & ext & 150 & 2.5 (15) & 3.0 & 1, 5\\
HNCO & 1.5 & 200 & 5.0 (16) & 5.0 & 1\\
N$_2$H$^+$ & ext & 50 & 5.0 (13) & 5.0 & 1\\
NH$_2$CHO & ext & 150 & 1.0 (14) & 3.0\\
SO & ext & 150 & 1.0 (15) & 5.0 & 1\\
SO$_2$ & 1.5 & 150 & 2.0 (17) & 3.0 & 4\\
\hline
\multicolumn{6}{c}{\textbf{G327.3-0.6}}\\
\hline
c-C$_2$H$_4$O & ext & 150 & 8.0 (14)&  5.0 & 1\\
C$_2$H$_5$CN & 1.5 & 200 & 3.0 (17) & 3.0\\
             & ext & 50 & 8.0 (15) & 3.0 \\
CH$_3$CN & 1.1 & 300 & 5.0 (17) & 4.0 & 3, 4\\
         & ext & 50 & 5 (14) & 4.0\\
CH$_3$OCH$_3$ & ext & 200 & 1.0 (16) & 3.0 & 2\\
CH$_3$OCHO & 1.5 & 150 & 2.0 (18) & 3.0 & 2\\
CH$_3$OH  & ext & 50 & 2.0 (17) & 5.0 & 1\\
HNCO & ext & 150 & 1.0 (15) & 5.0 & 1\\
N$_2$H$^+$ & ext & 50 & 3.0 (14) & 5.0 & 1\\
NH$_2$CHO & ext & 150 & 3.0 (14) & 5.0 & 2\\
SO$_2$ & 1.0 & 250 & 2.0 (18) & 8.0 & 4\\

\hline
\\
\multicolumn{6}{l}{(1) Based on one or a few lines only.}\\
\multicolumn{6}{l}{(2) Based on weak or partially blended lines only.}\\
\multicolumn{6}{l}{(3) Includes lines of the vibrationally excited state.}\\
\multicolumn{6}{l}{(4) Includes optically thick lines.}\\
\multicolumn{6}{l}{(5) Based on C$^{34}$S assuming $^{32}$S/$^{34}$S=23.}\\
  \end{tabular}
  \caption{Molecular parameters determined for \ngc\ and \gthree.
\new{LTE ``excitation temperature'',  $T_{\rm ex}$, column density, $N$(mol), and
linewidth,   $\Delta v$, were determined as described in the text.}
For $N$(mol), the notation $a(b)$ means $a\times 10^b$.
Footnotes
(1) or (2) imply that the determination of the molecular parameters is not
very reliable. The label \emph{ext} means extended emission.}
  \label{tab:model}
\end{table}

\section{Discussion}

We refrain from providing a list of all identified transitions in the spectra,
but discuss each species individually.  
Lower frequency data of \ngc\ were obtained
by \citet{nummelin98} [N98] and \citet{gibb00} [G00].  

\vspace*{-3ex}
\paragraph{c-C$_2$H$_4$O} Ethylene oxide was already found by
N98 in both sources. We detect only one line, with a lower energy level
corresponding to 170~K, thus tracing hotter gas than was traced by N98.
Indeed, we are unable to fit our one line with the low excitation temperatures
of 38~K or 27~K (for \ngc\ and \gthree, respectively) that N98 derive, so we
assume a temperature of 150~K to derive the column density. 

\vspace*{-3ex}
\paragraph{CH$_3$CN}

For methyl cyanide, we observe enough highly excited lines, including from the
vibrationally excited $v_8=1$ state, to determine source sizes and
temperatures quite reliably.  This is particularly remarkable in \gthree,
where the ground state lines become very optically thick, and point to a very
small, hot central source. 
The high temperature we
derive comes from the need to fit the $v_8=1$ lines, which are about 900~K
above ground, but the exact temperature is difficult to determine, due to the
high optical depth even in the $v_8=1$ state. The evidence for an extended,
cooler component in \gthree\ is weak, and is only suggested by an excess
emission in the low energy $K=0, 1, 2$ components. 

\vspace*{-3ex}
\paragraph{CH$_3$OCH$_3$}

Dimethyl ether displays in our frequency range transitions originating from
both low and high level energies, the latter particularly in the 810~GHz band
toward \ngc.  As as the case with methyl cyanide, the temperatures and source
sizes are quite well determined, and agree with N98, who also find it
necessary to fit two components. In \gthree, due to the absence of the 810~GHz
lines, the source can be modeled by one hot, optically thin
component. 

\vspace*{-3ex}
\paragraph{CH$_3$OCHO}

Methyl formate has mostly highly excited lines in the survey range, so the
temperature and size of the compact component are quite well determined in
\ngc.  N98 do not observe this \cut{line} \new{molecule}, so we cannot
compare. In \gthree, the lines tend to be weak and/or blended, so the
determination of column densities is not very accurate, but a compact, hot
component with many optically thick lines seems likely, in agreement with the
results from G00. 

\vspace*{-3ex}
\paragraph{CH$_3$OH}
Methanol displays a plethora of high-excitation lines at 820~GHz, but only one
fairly low-excitation line in the 460~GHz band.  Consequently, for \ngc, we
can establish the presence of a hot, compact component.  N98 also find a hot
component, but with a lower temperature than we do.  The reason is probably
that N98 only had access to lines coming from much lower energies than our
survey.  In \gthree, the 460~GHz line is only sensitive to the extended, less
excited gas, but does not exclude the existence of a hot, compact component,
which in this source has been found by G00.


\vspace*{-3ex}
\paragraph{HNCO}

In \ngc, HNCO is observed in the $K_a=0, 1, 2$ states of the $J=21-20$
transition.  The $K\ne 0$ lines are pumped by FIR radiation at 300 $\mu$m
($K=1$) and 110 $\mu$m ($K=2$), as shown by \citet{churchwell86}.  The lines
come from a compact, hot source. In \gthree, only the $K=1$ line is in the
band, making a determination of the temperature impossible.  





\vspace*{-3ex}
\paragraph{NH$_2$CHO}
Formamide is discovered with some weak, high-excitation lines, and has been
modeled as a hot component with no size restriction.  G00 find formamide to be
emitted by a moderately sized region with a moderate temperature. 



\vspace*{-3ex}
\paragraph{SO$_2$}
Sulphur dioxide displays lines originating from various, but including very
high, excitation levels, particularly traced by the 460~GHz lines.  The
emission can be modeled as arising from a single compact and hot component,
which gives rise to optically thick lines in both sources.  In \gthree, this
finding agrees well with G00.  In Orion, emission from the sulphur-bearing
molecules is dominated by the plateau (outflow) component, but highly excited
and vibrationally excited SO and \sotw\ molecules do show characteristics of
the hot core.

\vspace*{-3ex}
\paragraph{U-lines}
The number of \new{unidentified (U)}-lines we find is low, about 7 out of 245
lines. This is partly due to the fact that, although the survey is quite
sensitive, we are not yet confusion-limited in most bands -- most unidentified
lines are in the lower intensity range.  Identification is made difficult by
the large frequency errors for many species in the catalogs.  

\section{Conclusions}
In this \emph{Letter}, we have reported the first attempt to perform a
submillimeter line survey \revised{shortward of 800 $\mu$m} in a high mass
star forming region other than Orion-KL. The results are intriguing: not only
is it possible to do these surveys in regions which are between 3 and 6 times
more distant than Orion, with similar signal-to-noise ratio, but one does see
a higher line density too.  Submillimeter observations, since many lines
originate from highly excited levels, allow a glimpse into the very heart of
the star formation.  More systematic, and even higher-frequency observations
with Herschel/HIFI, and particularly spatially resolved observations with ALMA
will allow a much more detailed picture of these regions, and will determine
the shape of the exact surroundings of newly born high-mass stars.  Until
then, surveys such as the one presented here will be invaluable to study these
objects.

\begin{acknowledgements}
  We are grateful to the APEX commissioning team.  Glenn White very kindly
  made his line survey available in electronic form.  Line survey work would
  not be possible without the tireless work of the maintainers of molecular
  databases, particularly CDMS and JPL, and by the numerous spectroscopists
  providing the input.

\end{acknowledgements}


\end{document}